\documentclass[prl, 11pt, aps, showpacs, longbibliography,preprint,]{revtex4-2}

\usepackage{amssymb,bm}
\usepackage{graphicx}
\usepackage{color}
\usepackage{amsmath} 
\usepackage{setspace}
\usepackage{enumerate}
\usepackage{bbold}
\usepackage{esint}
\usepackage{euscript}
\usepackage[bookmarks, colorlinks=true, breaklinks]{hyperref}
\hypersetup{linkcolor=blue,citecolor=blue,filecolor=black,urlcolor=blue}

\newcommand{\bra}{\left< }
\newcommand{\ket}{\right>}

\newcommand{\norm}[1]{\left| {#1} \right|}

\begin{document}

\title{Lamellar fluctuations melt ferroelectricity}

\author{G. G. Guzm\'{a}n-Verri\footnote{gian.guzman@ucr.ac.cr}}
\affiliation{Centro de Investigaci\'{o}n en Ciencia e Ingenier\'{i}a de Materiales, Universidad de Costa Rica, San Jos\'{e}, Costa Rica 11501,}
\affiliation{Escuela de F\'{i}sica, Universidad de Costa Rica, San Jos\'{e}, Costa Rica 11501,}
\affiliation{Department of Materials Science and Metallurgy, University of Cambridge, Cambridge, UK CB3 0DS.} 
\author{C. H. Liang}
\affiliation{James Franck Institute, University of Chicago, Chicago, Illinois, USA 60637,}

\affiliation{Pritzker School of Molecular Engineering, University of Chicago, Chicago, Illinois, USA 60637.}
\author{P. B. Littlewood\footnote{littlewood@uchicago.edu}}
\affiliation{James Franck Institute, University of Chicago, Chicago, Illinois, USA 60637.}
\date{\today}

\begin{abstract}
We consider a standard Ginzburg-Landau model of a ferroelectric whose electrical polarization is coupled to gradients of elastic strain~\cite{Zubko2013a, Yudin2013a, Tagantsev2016a, Wang2019a}. At the harmonic level, such flexoelectric interaction is known to hybridize acoustic and optic phonon modes and lead to phases with modulated lattice structures that precede the symmetry broken state for sufficiently large couplings~\cite{Vaks1968a, Axe1970a, Aslanyan1980a, Heine1981a, Yudin2014a, Morozovska2016a}. Here, we use the we use the self-consistent phonon approximation to calculate the effects of thermal and quantum polarization fluctuations on the bare modes to show that such long-range modulated order is unstable at all temperatures. 
We discuss the implications for the nearly ferroelectric SrTiO$_3$ and KTaO$_3$, and propose that these systems are melted versions of an underlying modulated state which is dominated by non-zero momentum thermal fluctuations except at the very lowest temperatures.
\end{abstract}

\maketitle

A ferroelectric (FE) material is defined by spontaneously broken inversion symmetry, usually due to a small and coherent displacement of atoms within the unit cell and hence a spontaneous electrical polarization ${\bm P}$~\cite{Strukov1998a}. Because lattice displacements generate the order, the homogeneity of the polarization depends on coupling between neighboring unit cells, and over longer distances by a continuous elastic strain $\epsilon$.  In a cubic (or other high symmetry FE) the principal order parameter (OP) ${\bm P}$ belongs to a different irreducible representation from the strain fields $\epsilon$ at zero wavevector (${\bm q}=0$). This means that the leading order electrostrictive coupling in a uniform Landau theory for the free energy is bilinear in polarization and linear in strain (i.e. $\mathcal{O}\left(\epsilon P^2\right)$. However, also there will be in general a {\it flexoelectric} coupling~\cite{Zubko2013a, Yudin2013a, Tagantsev2016a, Wang2019a} (i.e. $\mathcal{O}\left(P \nabla \epsilon \right)$), and it is well known that such couplings may give rise to modulated incommensurate phases within a harmonic theory, provided that the coupling is big enough~\cite{Vaks1968a, Axe1970a, Heine1981a, Aslanyan1980a, Yudin2014a, Morozovska2016a}. Note that the flexoelectric coupling - being harmonic - is in principle more relevant than the nonlinear electrostrictive coupling; flexoelectricity will be the focus of this paper. 

The materials SrTiO$_3$ (STO) and KTaO$_3$ (KTO) have long been cited as examples of quantum paraelectrics (QPEs). In both cases the dielectric constant follows a Curie-Weiss (CW) law divergence with a finite-temperature intercept, yet at low temperatures saturates at extremely large values~\cite{Muller1979a}  and there is no FE phase transition. The idea of quantum tunnelling of domains that in a mean-field like (and massive) system have already grown to be very large before quantum effects take over is disconcerting and has, over the years, generated considerable interest in the general theory of a FE quantum critical point ~\cite{Muller1991a, Martonak1991a, Tosatti1994a, Martonak1994a, Zhong1996a, Roussev2003a,  Bussmann-Holder2007a,Palova2009, Conduit2010a}.

Recent experiments, however, have revealed phase diagrams and lattice dynamics in STO and KTO which cannot be explained with our current understanding of these materials~\cite{Chandra2017a}. Specifically, high precision measurements of the dielectric response have found several crossovers (specially seen under pressure)~\cite{Coak2018a, Coak2020a, Coak2019b, Rowley2014a}, and, very significantly, inelastic neutron scattering experiments observe softening of a transverse acoustic (TA) mode at a small but finite wavector in STO ($\simeq 0.025\,$rlu)~\cite{Fauque2022a} and at a larger one in KTO ($\simeq 0.1\,$rlu)~\cite{Axe1970a}. The latter results are consistent with momentum-resolved electron energy-loss spectroscopy~\cite{Kengle2022a} and a previous neutron scattering study~\cite{Vacher1992a} which show an unusual softening of the TA branch, and with a first-principle simulation~\footnote{See Fig. 2 in Supplementary of Ref.~\cite{He2020a}} which found a dip in the acoustic dispersion.

Soft acoustic phonons with a minimum at a finite wavector are characteristic of systems approaching a structural instability corresponding to the onset of a long-wavelength modulation~\cite{Blinc1986a, Scott1983a}, and can be understood as the consequence of level-repelling hybrid modes associated with a sufficiently large coupling between a primary OP (e.g. polarization) and gradients of a secondary OP (e.g. strain)~\cite{Heine1984a}. However, harmonic and mean-field solutions of such models~\cite{Vaks1968a, Axe1970a, Heine1981a, Aslanyan1980a, Yudin2014a, Morozovska2016a} predict the condensation of the low energy branch at a finite wavevector ${\bm q}_{mod}$, thus leading to a modulated phase which is not observed in STO and KTO. Nor can these solutions generate crossovers. 

In this letter, we invoke a well-known model of flexoelectricity~\cite{Zubko2013a, Yudin2013a, Tagantsev2016a, Wang2019a} and, crucially, solve it within the self-consistent phonon approximation (SCPA) in order to account for thermal and quantum fluctuations of polarization of the disordered phase. We show that in the presence of purely thermal fluctuations, the phase space for orientational fluctuations is large enough that the modulated transition is suppressed to zero temperature, and that by including zero-point quantum fluctuations, the correlation length will remain finite even at absolute zero.
\begin{figure}[htp!]
    \centering
    \includegraphics[scale=0.38]{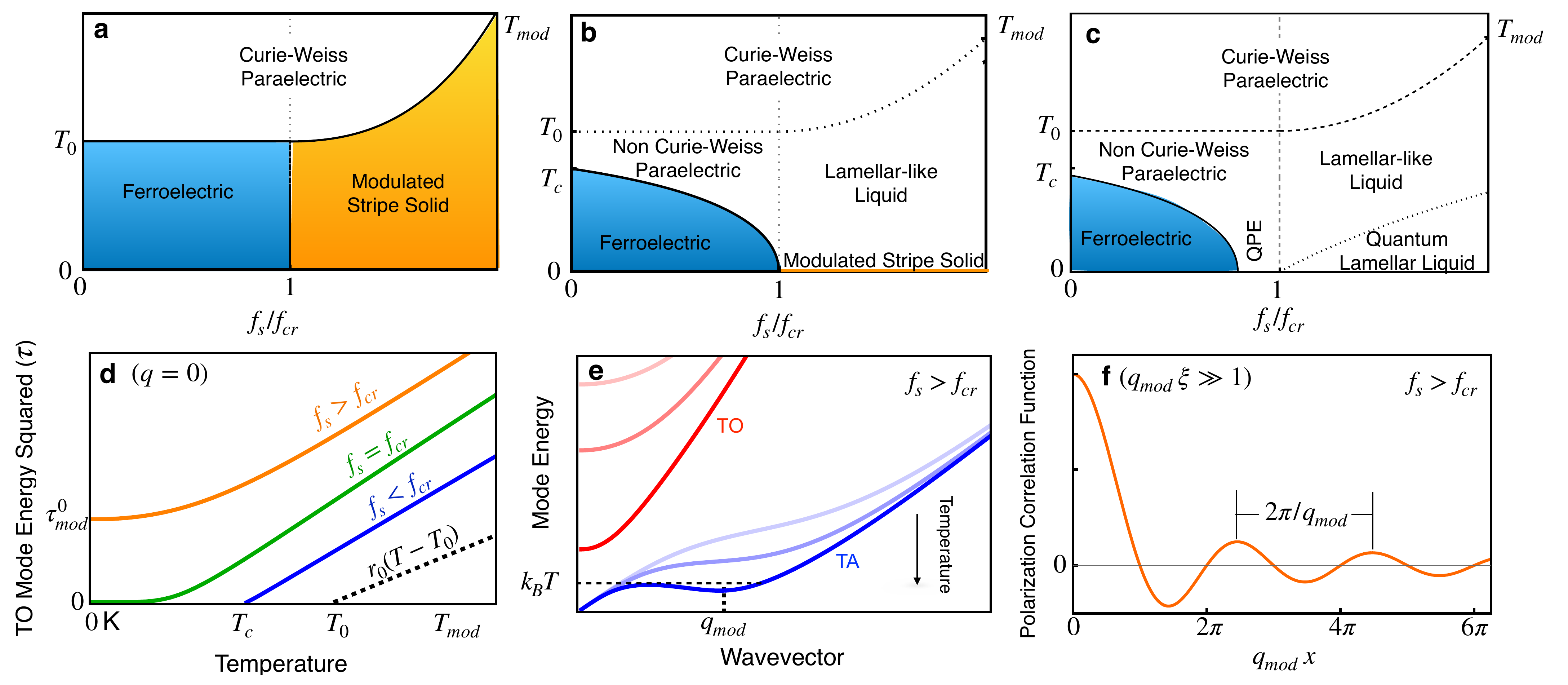}
    \caption{Phase boundaries in the {\bf (a)} harmonic approximation and in the SCPA {\bf (b)} ignoring and {\bf (c)} including quantum fluctuations of polarization.   {\bf (d)} SCPA schematics of the temperature dependence of the zone-centre hybrid TO mode and the {\bf (e)} hybrid TA and TO phonon dispersions in the disordered phase. {\bf (f)} Schematic correlation function of polarization in the lamellar-like liquid with correlation length $\xi$ (see Eq.~(\ref{eq:polcorrelation})). QPE=quantum paraelectric.}
    \label{fig:phasediagram1}
\end{figure}
We estimate the parameters in comparison to KTO and STO, and find reasonable agreement with a model where the flexo-electric coupling is close to the classical critical value for the onset of a modulated phase. Our analysis suggests that the putative QPE phase of STO and KTO is largely explained by classical thermal modulated fluctuations, and that there is a second crossover to a regime dominated by quantum fluctuations at much lower temperature. However, the quantum regime in our theory does not require macroscopic quantum tunnelling, and instead arises more trivially due to zero-point occupation of orientational zero-modes.  In the light of these results, we propose that STO and KTO are incipient modulated dielectrics with incommensurate order caused by the coupling of polarization and strain inhomogeneity. In the underlying ordered phase, the rotational symmetry of the crystal is broken as well as its translational symmetry, but only in one direction (${\bm q}_{mod}$). This would lead to intertwined stripe order parameters where the polarization and elastic strain are periodic with a wavevector ${\bm q}_{mod}$~\cite{Blinc1986a}. Such stratified ordering resembles the structural organisation  of the ``lamellar'' or ``smectic'' phases in liquid crystals: a nearly periodic array of parallel layers of polarizable molecules which do not possess long-range positional order within a layer~\cite{Chaikin1995a}. In a smectic the director vector ${\bm n}$ points orthogonal to the planar lamellae, as does here the wavevector ${\bm q}_{mod}$.  The underlying materials science is of course quite different.

Lastly, we comment on lacunae in the model, the most important of which is the neglect of crystal anisotropy and compositional disorder. We expect anisotropy to lead to an ordered (striped) phase at low enough temperatures; disorder to a glassy frozen version of the same. Both of these effects can suppress the quantum regime. We also suggest experiments to further test our proposed picture. 

We consider an isotropic polarizable, elastic medium with a polarization order parameter $P_\alpha \left( {\bm x} \right)$ and a linear strain $\epsilon_{\alpha \beta}({\bm x}) =  (1/2)  \left( \partial u_\alpha / \partial x_\beta +  \partial u_\beta / \partial x_{\alpha} \right)$ as a secondary order parameter in which 
$u_\alpha({\bm x})$ is the displacement field due to long-wavelength acoustic phonons ($\alpha,\beta=1,2,3$).
$P_\alpha \left( {\bm x} \right)$ is associated with a soft transverse optic (TO) mode, the condensation of which leads to the FE transition. This leads us to consider the GL Hamiltonian that is a sum of three terms: for the polarization $H_\text{pol}$; the elastic modes $H_\text{elastic}$; and the static flexoelectric coupling $ H_\text{flexo}$\,\cite{Yudin2013a},
\begin{align}
	\label{eq:Hr}
		H  = H_\text{pol}  + H_\text{elastic}  + H_\text{flexo},
\end{align}
where,
\begin{align*}
H_\text{pol}  & = \frac{1}{2}\int d^3 x d^3 x^\prime  \sum_{\alpha, \beta} U_{\alpha \beta}({\bm x},{\bm x}^\prime) P_{\alpha} \left( {\bm x} \right)  P_{\beta} \left( {\bm x}^\prime \right)
+
u \int d^3 x \sum_{\alpha, \beta}   P_\alpha^2  \left( {\bm x} \right) P_\beta^2 \left( {\bm x} \right),
\end{align*}
\begin{align*}
H_\text{elastic}  & = \frac{1}{2}\int d^3 x  \sum_{\alpha, \beta, \gamma, \lambda} C_{\alpha \beta \gamma \lambda} \, \epsilon_{\alpha \beta}({\bm x})
 \epsilon_{\gamma \lambda}({\bm x}),
\end{align*}
and,
\begin{align*}
    H_\text{flexo} = \frac{1}{2} \int d^3 x  \sum_{\alpha \beta \gamma \lambda}  f_{\alpha \beta \gamma \lambda} \, \left[ \epsilon_{\beta \gamma}({\bm x}) \frac{\partial P_\alpha ({\bm x})}{\partial x_\lambda} - P_\alpha ({\bm x}) \frac{\partial \epsilon_{\beta \gamma} ({\bm x})}{\partial x_\lambda} \right].
\end{align*}
Here, $U_{\alpha \beta}({\bm x},{\bm x}^\prime)=r \delta_{\alpha \beta} \delta({\bm x}-{\bm x}^\prime) + F_{\alpha \beta}({\bm x},{\bm x}^\prime)$ is the bare propagator  of the optic phonons, $F_{\alpha \beta}({\bm x},{\bm x}^\prime)$ is the dipole tensor with Fourier transform
$F_{\alpha \beta} \left( {\bm q} \right) = c q^2 \delta_{\alpha \beta} + g_0 \left( q_{\alpha} q_{\beta} / q^2 \right) - h_0 q_{\alpha} q_{\beta}$~\cite{Aharony1973a}.  $C_{\alpha \beta \gamma \delta}$ is the elastic constant tensor in units of energy per unit volume, $f_{\alpha \beta \gamma \lambda}$ is the flexocoupling tensor, and $r=r_0(T-T_0)$ where $T_0$ is the mean field transition temperature to a homogeneous FE state at $f_{\alpha \beta \gamma \lambda}=0$.

In the absence of flexoelectric coupling 
and non-linearities $(u=0)$, the phonon excitations of $H$ are well-known: there are two doubly-fold degenerate TO and TA modes with (squared) frequencies  $\omega_\text{TO}^2({\bm q}) = r + c q^2$ and   $\omega_\text{TA}^2({\bm q}) = (\hbar^2 C_s / \rho)  q^2$, respectively, and 
two singly-degenerate longitudinal optic  and acoustic modes with (squared) frequencies $\omega_\text{LO}^2({\bm q}) = r +g_0 + (c +h_0) q^2$ and
$\omega_\text{LA}^2({\bm q}) = (\hbar^2( C_v + (4/3)C_s) / \rho)  q^2$, respectively.  $C_v$ and $C_s$ are the bulk and shear elastic moduli, respectively, and $\rho$ is the density of the solid. 

In this paper, we focus on the effects on the hybrid transverse modes only as the level repulsion is significantly weaker in the longitudinal branches due to the characteristically large depolarizing fields of FEs which gap the longitudinal optic excitations. Moreover, it is shown in the Supplementary Note 1 (SN1) that in isotropic media the flexoelectric interaction does not mix the transverse and longitudinal excitations.

We now discuss our results.  Figure~\ref{fig:phasediagram1}\,(a) shows schematics of the phase boundary (PB) computed from the condensation of the bare modes (i.e., $u=0$) in the parent phase (see SN1). For small shears of the flexocoupling tensor (i.e., $f_s < f_{cr} = \sqrt{c\, C_s}$), the soft zone-center TO mode  condenses at a transition temperature which is independent of $f_s$ and equal to $T_0$, thus leading to a FE transition.   For large flexoelectric shears (i.e., $f_s > f_{cr}$), a soft minimum develops at a non-zero wave-vector ${\bm q}_{mod}$ of the TA branch, which upon condensation at a temperature $T_{mod}>T_0$, would lead to the modulated phase with coupled strain and polarization order parameters that are periodic with a wavevector ${\bm q}_{mod}$~\cite{Blinc1986a}. The zone-center TO mode remains gaped with energy squared $\tau^0_{mod} \equiv r_0(T_{mod}-T_0)$. At  $f_s=f_{cr}$, ${\bm q}_{mod}=0$ and both the TO and TA modes condense $T_{mod} = T_0$.

We now assess the stability of the bare PB. Near above $T_0$ and $T_{mod}$ and in the classical limit,  the local correlation functions of polarization are approximately given as follows (see SN2),
\begin{align*}
    \left< {\bm P}^2  \right>_0  \propto \begin{cases}
    1 - b \left(T-T_0\right)^{1/2}, &   f_s < f_{cr}, \\
    \left( T-T_0 \right)^{-1/4}, & f_s = f_{cr},\\
    \left( T-T_{mod} \right)^{-1/2}, & f_s > f_{cr},
    \end{cases}
\end{align*}
where $ \left< ...\right>_0 $ denotes thermal average at the Gaussian level and $b$ is a constant. The PB is thus strongly modified by the inclusion of fluctuations: while the FE transition prevails at $T_0$, the long-range modulated phase change is unstable, except at absolute zero where $\left< {\bm P}^2  \right>_0 =0$.

We now consider non-linearities (i.e., $u>0$) 
in the SCPA (see SN3). We first discuss the impact of thermal fluctuations alone and then introduce quantum effects. Fig.~\ref{fig:phasediagram1}\,(b) shows the 
 purely classical PB. For $f_s < f_{cr}$, the coupling to finite momentum elastic fluctuations suppresses the FE transition temperature and leads to a thermal regime where the squared of the soft zone-center TO phonon frequency $\tau$ (or equivalently the inverse dielectric constant) separates from the CW law behavior, as shown in Fig.~\ref{fig:phasediagram1}\,(d). 
For $f_s > f_{cr}$, the bare PB turns into a crossover. This is illustrated best by the SCPA dispersions shown in Fig.~\ref{fig:phasediagram1}\,(d). As mentioned above, the bare  coupled acoustic mode softens to zero at a non-zero momentum ${\bm q}_{mod}$. However, in an isotropic theory, such as ours, only the modulus of ${\bm q}_{mod}$ is determined but not the direction. The phase space for fluctuations in the {\em direction} of ${\bm q}_{mod}$ lies on the surface of a sphere of one dimension less than the physical dimension. This all but guarantees that thermal fluctuations {\it at any finite temperature} will disorder the direction, so there will be no long-range stripe order. This is very analogous to fluctuating laminar or nematic soft matter systems~\cite{Chaikin1995a}. Consequently, the harmonic transition in Fig.~\ref{fig:phasediagram1}\,(a) becomes a crossover in Fig.~\ref{fig:phasediagram1}\,(b), and the  acoustic excitations in the vicinity of ${\bm q}_{mod}$ acquire a gap, as seen in Fig.~\ref{fig:phasediagram1}\,(e). 
The polarization correlations of the melted phase are sinusoidal functions attenuated by an exponentially decaying envelope with a correlation length $\xi \propto \left( \tau-\tau^0_{mod} \right)^{-1/4}$ longer than the period of the modulation, i.e.,
\begin{align}
    \bra {\bm P}({\bm x}) \cdot {\bm P}({\bm 0})  \ket \sim e^{-x/(2\,  q_{mod} \, \xi^2)} \,  \sin \left( q_{mod} \, x \right)/\left( q_{mod} \, x \right),~~~~~~(q_{mod}\, \xi \gg 1),
    \label{eq:polcorrelation}
\end{align}
where $x=\norm{\bm x}$ and $q_{mod}=\norm{{\bm q}_{mod}}$ (see SN3). This is schematically shown in Fig.~\ref{fig:phasediagram1}\,(f). Upon cooling, $\xi$ would grow as $\tau$ approaches $\tau_{mod}^0$ (Fig.~\ref{fig:phasediagram1}\,(d)) and diverge at absolute zero, thus restoring the long-range stripe order, as shown in Fig.~\ref{fig:phasediagram1}\,(b). 

\begin{figure}[htp!]
    \centering
    \includegraphics[scale=1]{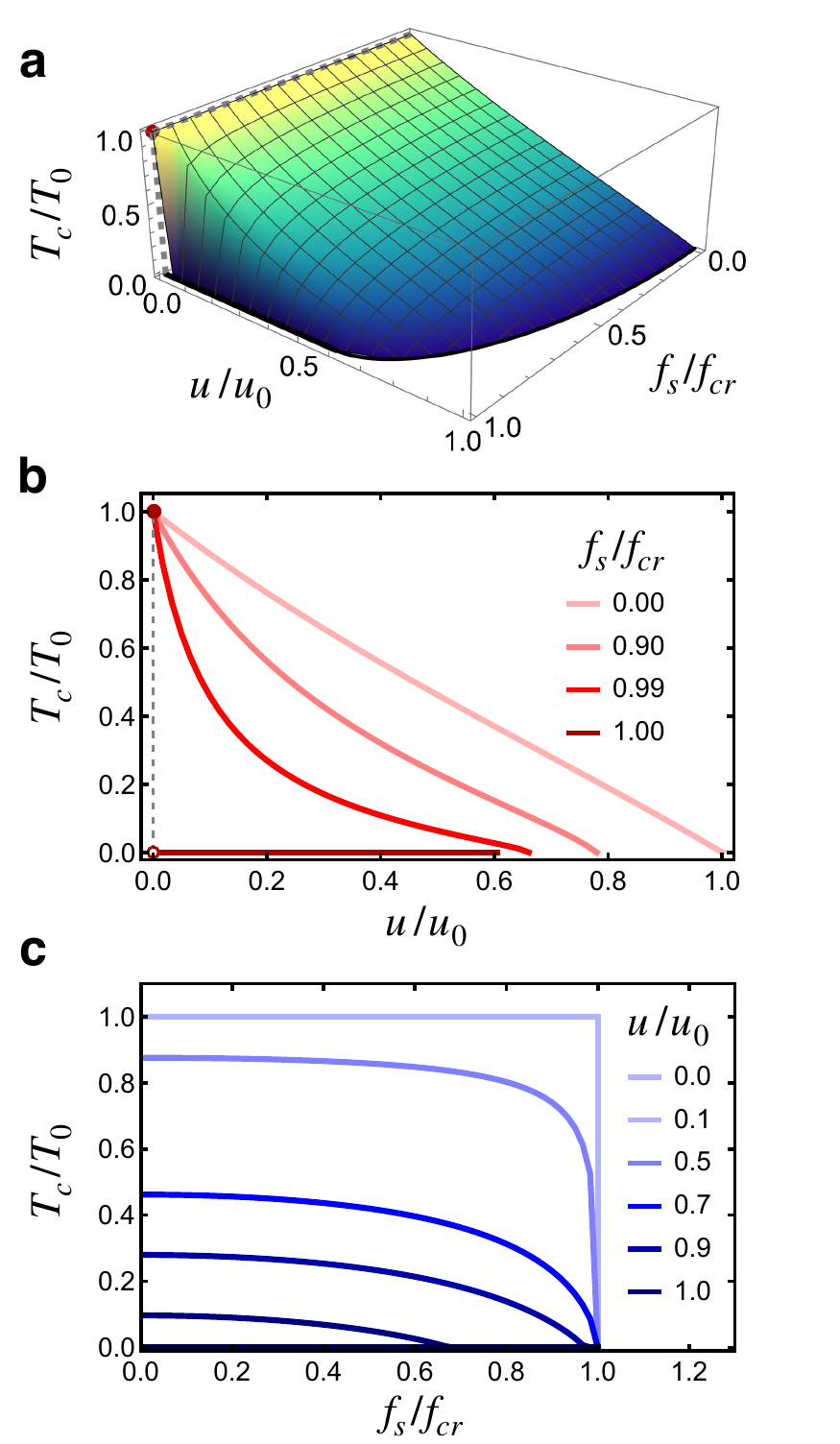}
    \caption{{\bf (a)} Dependence of the ferroelectric transition temperature $T_c$ on the flexoelectric shear coupling constant $f_s$ and the quartic anharmonic coefficient $u$ of the polarization. 
    {\bf (b)}-{\bf (c)} Cross-sections of {\bf (a)} in the $T_c-u$ and $T_c-f_s$ plane, respectively. 
    $u_0$ is the value of $u$ at $T_c=0\,$K and $f_s=0$.}
    \label{fig:phasediagram2}
\end{figure}

We now asses the effects of quantum polarization fluctuations. In our model, the nonlinearity of the theory is controlled by the quartic coefficient $u$, as shown in Fig.~\ref{fig:phasediagram2}. For $f_s=0$, the FE transition temperature drops approximately linearly in $u$, vanishing at a quantum critical point. For $0<f_s<f_{cr}$, the critical nonlinearity is reduced, and there is a much more substantial parameter regime where quantum fluctuations alone destroy the transition and lead to the standard QPE phase, as it is shown in  Fig.~\ref{fig:phasediagram1}~(c). This is consistent with the much larger phase space for orientational fluctuations when the long-range modulated order would be at finite $q$. 

For $f_s>f_{cr}$, the zero-point fluctuations keep the energy gap at ${\bm q}_{mod}$ open even at absolute zero, thus destabilising the classical $T=0\,$K transition shown in Fig.~\ref{fig:phasediagram1}\,(b). 
In addition, the temperature at which the quantum fluctuations become relevant marks another crossover in the spectrum, where the correlation length will saturate. This occurs when $k_B T$ is comparable to the energy of the excitations around ${\bm q}_{mod}$, as shown in Fig.~\ref{fig:phasediagram1}\,(e). So, for $f_s > f_{cr}$ we predict that there will be two crossovers as $T$ is lowered, and no long-range stripe order,
as shown in Fig.~\ref{fig:phasediagram1}\,(c). 
We stress that when zero-point quantum fluctuations are included, the only ordered phase that we find is ferroelectricity. But outside the envelope shown in Figure~\ref{fig:phasediagram2}\,(a) and for $f_s>f_{cr}$, we expect the lamellar-like fluctuations to have a very long-range correlation.

\begin{figure}[htp!]
    \centering
    \includegraphics[scale=0.65]{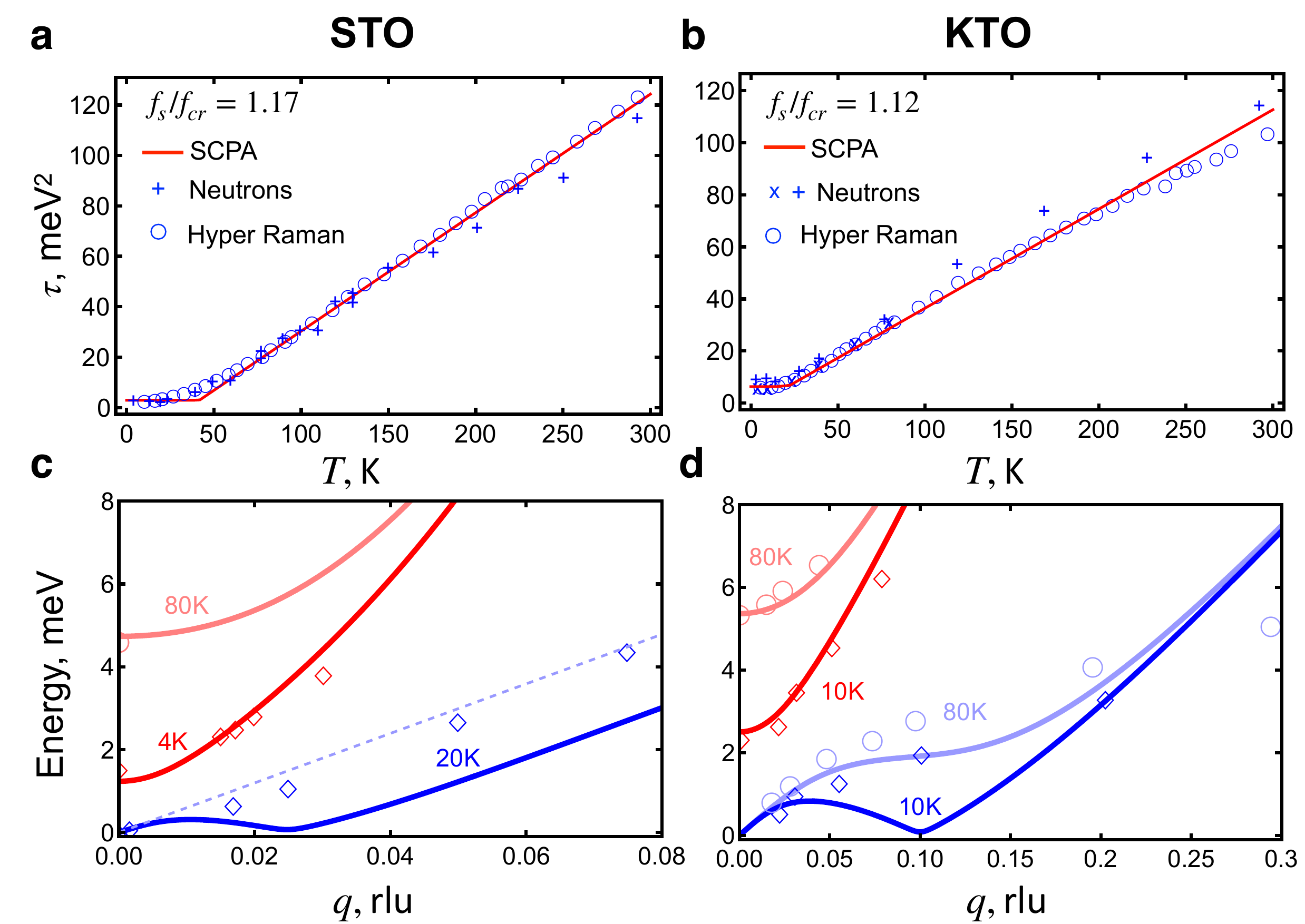}
    \caption{{\bf (a),(b)} Calculated temperature dependence of the zone center hybrid TO mode compared to neutron~\cite{Shirane1967a, Yamada1969a, Farhi2000a} and hyper-Raman~\cite{Vogt1995a} scattering experiments . {\bf (c),(d)} Calculated dispersion of the hybrid TA and TO phonons (solid lines) compared to neutron scattering experiments (circles and diamonds)~\cite{Fauque2022a, Farhi2000a}. Dashed line in (d) is the bare linear TA dispersion for $f_s=0$ with a sound velocity of 4800\,m/s (as shown in Ref.~\cite{Fauque2022a} for comparison). Model parameters are given in Table~\ref{t:parameters}.}
    \label{fig:Comparison}
\end{figure}

\begin{table}[htp]
\caption{Model parameters.} 
   \begin{tabular}{l|r|r} 
   \hline
    &  \hspace{1cm} KTO &   \hspace{1cm} STO  \\ \hline \hline 
       $r_0\,$[meV$^2$ K$^{-1}$]  & $0.38$\footnotemark[2] & $0.47$\footnotemark[5]    \\
       $T_0\,$[K]  & $4.54$\footnotemark[2]  & $35.5$\footnotemark[5]   \\
       $u\,$ [$10^{-4}$ meV$^3$   rlu$^{-3}$]  & $7.19$\footnotemark[3]   & $7.61$\footnotemark[5]   \\
       $c\,$[$10^3$\, meV$^2$ rlu$^{-2}$]  & $4.828$\footnotemark[4]  &     $13.5$ \\
       $a_t$\footnotemark[1]\, [$10^3\,$ meV$^2$  rlu$^{-2}$]  & $1.553$\footnotemark[4] &  $2.739$\footnotemark[6]   \\
     $v_t$\footnotemark[1]\, [$10^3\,$ meV$^2$ rlu$^{-2}$] &  $3.067$  & $6.737$  \\
     $h\,$ [$10^{4}$ meV$^4$  rlu$^{-4}$]  & $6.15$  & $406$  \\
       \hline 
      $v_t/v_{cr} = f_s/f_{cr}$   & $1.12$    & $1.17$  \\ \hline 
   \end{tabular}
    \label{t:parameters}
    \footnotetext[1]{$a_t = \hbar^2 C_s / \rho $ and $v_t  = \hbar  f_s / \sqrt{\rho}$.}
    \footnotetext[2]{From Ref.~\cite{Fujishita2016a}.}
    \footnotetext[3]{From Ref.~\cite{Skoromets2016a}.}
   \footnotetext[4]{From Ref.~\cite{Farhi2000a}.}
\footnotetext[5]{From Ref.~\cite{ModernPerspective2007a}.}
\footnotetext[6]{From Ref.~\cite{Carpenter2007a}.}
\end{table} 

We now turn to comparisons to STO and KTO. 
Table~\ref{t:parameters} gives the values for the model parameters, which for the most part were taken from the literature while others were obtained from experimental data, e.g., $f_s$ was fitted to the reported TA anomalies ($0.025\,$rlu for STO~\cite{Fauque2022a}, and $0.1\,$rlu for KTO~\cite{Axe1970a}). The details can be found in SN4. For both materials, we find agreement between the calculated and measured 
temperature dependence of the TO frequency, as shown in Figs.~\ref{fig:Comparison}\,(a) and ~\ref{fig:Comparison}\,(b). We note the apparent kink in the calculated frequencies is a consequence of the small values of $u$ in our paremetrization. As shown in SN3, $\tau$ is a smooth function of $T$ at all temperatures. Figures~\ref{fig:Comparison}\,(c) and ~\ref{fig:Comparison}\,(d) show the calculated and observed phonon dispersion curves.  While our model and parametrization clearly overestimate the repulsion between the TA and TO branches, it generates a minimum at a finite $q$ in the hybrid acoustic dispersion while preventing symmetry breaking down to the lowest temperatures, as observed in experiments~\cite{Fauque2022a, Axe1970a}. We stress that no such behavior occurs for $f_s \leq f_{cr}$, nor could we obtained a physically reasonable parametrization in this regime, as explained in SN4. We attribute the quantitative discrepancies to ignoring crystal anisotropy (e.g. low temperature STO is tetragonal while KTO is cubic),  broadening,  non-zero lifetimes, electrostriction, and coupling to other lattice modes (e.g., rotations of the TiO$_6$ octahedra).
Our fits place both materials in the liquid phase of Fig.~\ref{fig:phasediagram1}\,(c) ($f_s/f_{cr}=1.17$ for STO and $f_s/f_{cr}=1.12$ for KTO). Of course, these values should be taken as rough estimates, as our simplified model, approximate solution, and currently available experimental data prevent us from determining $f_s$ precisely. Nonetheless, they show that both materials are near a modulated instability. 

This point of course connects to a long debate about the nature of the FE phase transition as ``displacive'' or ``order-disorder''  \cite{Jona1993, Comes1970}. Polar nanodomains are clearly seen at room temperature in strained (and therefore low-temperature ferrolectric) films of STO though their presence has been attributed to long-range Coulomb forces rather than strain \cite{Salmani2020}. Nuclear magnetic resonance (NMR) is a technique that is sensitive to local charge disproportionation and electric field gradients, and was useful in identifying charge-density waves in transition metal dichalchogenides, even when not fully ordered~\cite{Stiles1976, Pfeiffer1982}.  NMR measurements on STO~\cite{Zalar2005} have been interpreted in terms of dynamic fluctuations of the Ti-ion into off-center sites coupled in a biased way to elongations of the unit cell; however the length scale of this correlation could not be determined and the measurements did not go below $25\,$K. Data at a lower temperature would be helpful. Ab initio calculations of STO\,\cite{Shin2021}  concluded that quantum lattice fluctuations are necessary to stabilise the paraelectric phase - but spatially modulated phases were not considered in that work. Modulated phases could also be in principle seen in ab initio calculations, albeit using very large supercells. In that context we note  recent calculations \cite{Zhao2021a, Zhao2022a} which predict that STO and cubic BaTiO$_3$ should be polymorphic at $T=0\,$K, with a disordered arrangement of off-site Ti distortions. A disordered ground state at absolute zero violates the third law, and perhaps is associated instead with the failure to match the supercell to the natural incommensurate period.
 
 We now discuss our simplifying approximations.  We have ignored crystal anisotropy, in which case the orientational modes will have a preferred direction. Once the temperature becomes considerably smaller than the anisotropy in the spectrum, the phase space volume of critical fluctuations will be reduced, which may lead to the stabilization of an ordered modulated phase which could precede the establishment of an ordered FE phase (this is of course the standard situation seen, for example, in quartz~\cite{Aslanyan1978a}). Similarly, disorder will break the $\mathcal{O}(3)$ symmetry and should be expected to lead to a glassy phase where orientational fluctuations are frozen. This is a familiar situation in, e.g. charge-density-wave  systems, where impurities and disorder pin its order parameter and generate a strongly nonlinear response to applied electric fields~\cite{Fukuyama1978a, Lee1979a}.  
 Our conclusion that STO should be near a modulated instability is in stark contrast with ab initio calculations\,\cite{Stengel2016a}, which suggest that $f_s/ f_{cr} \simeq 0.7$ at most.  We note, however, that we are using a phenomenological continuum theory where the parameters are determined from experiment (but nonetheless can make explicit predictions because the number of parameters is small). We cannot assert that the parameters can be directly calculated from harmonic microscopic theory because we have already coarse-grained to a scale of many unit cells, so that, for example, twin boundaries and defect structures have been integrated over \cite{Pesquera2018a}. Therefore, the difference between parameters obtained from macroscopic fits and those determined by microscopic calculations hints at some mesoscale physics to be unearthed.

Our self-consistent harmonic solution of the model implicitly assumes that in the melted phases the amplitudes of fluctuations are small. Beyond this regime, electrostrictive couplings (effectively quartic in the primary order parameter) will be important. These of course enhance the anisotropy (at least at very long length scales) but also effectively soften the sound velocity and hence enhance the flexoelectric effect. In a strong coupling theory, the model would become one of quasi-periodic FE domains separated by fluctuating, sharp domain walls. In a better theory, the modes will acquire a linewidth, as seen in experiment. This is an interesting matter for future exploration, but we expect that the qualitative features of our theory should survive.

We conclude that a strong enough flexoelectric coupling 
and purely thermal fluctuations of polarization can mimic quantum paraelectricity by saturating the low-temperature dielectric constant.
We explain that in an isotropic system, the divergent orientational fluctuations will suppress ordering even at low temperatures so that one should expect a series of crossovers upon cooling: from a high-temperature CW law; to a classical fluctuating lamellar phase; to a quantum lamellar phase; and finally to a striped phase when crystal anisotropy becomes important. Such a sequence of crossovers is loosely consistent with experiments in STO \cite{Coak2018a, Coak2020a} and the soft finite $q$ modes have also been observed \cite{Fauque2022a,Kengle2022a}. Coak {\em et al.}\cite{Coak2018a} ventured that a loss peak seen in their data might be ascribed to quantum fluctuations of domain walls.
We  also stress that the quantum regime of our theory does not require tunneling of heavy atoms through potential barriers, but is simply zero-point physics in a quasi-harmonic well. With an albeit simplified theory, our parameters are physically reasonable and we believe our model provides a fair description of the disordered state in STO and KTO. 

In addition to the NMR experiments suggested above to further test our proposed scenario, we point out that measurements of the dynamical structure factor $S({\bm q},\omega)$ would detect fluctuations associated with the melting of the modulated phase. $S({\bm q},\omega)$ has of course been measured by neutrons and X-rays in  STO~\cite{Yamada1969a, Vacher1992a, He2020a} and KTO~\cite{Axe1970a,Farhi2000a}, but more data is needed in order to clearly resolve any dips along their acoustic branches. Finally, we point out that our model could be adapted to investigate flexoelectricity in other systems such as recently synthesized complex oxide membranes~\cite{Harbola2021a}, has a relation to structural Leggett modes in pyrochlore relaxors~\cite{Meier2021a}, and could be relevant to metallicity and superconductivity in doped STO~\cite{Rischau20107a}. 

\begin{acknowledgements}  
We acknowledge helpful feedback from Beno\^{i}t Fauqué, Massimiliano Stengel, Peter Abbamonte, Caitlin Kengle, Michael Norman, Xing He, and Siddharth Saxena. C.H.L. acknowledges support from NSF GRFP DGE-1746045, NSF INTERN, and DOE SCGSR. G.G.G.-V. acknowledges support from the Vice-rectory for Research at the University of Costa Rica under project No. 816-C1-601.
\end{acknowledgements}

\newpage

%\bibliography{references}

%

\clearpage
\pagebreak
\widetext
	\begin{center}
		\textbf{\large Supplementary Information}
	\end{center}
\vspace{0.5cm}
	\begin{center}
		G. G. Guzm\'{a}n-Verri$^{1,2,3}$, C. H. Liang$^{4,5}$, P. B. Littlewood$^4$ \\
		$^{1}${\it Centro de Investigaci\'{o}n en Ciencia e Ingenier\'{i}a de Materiales~(CICIMA),\\ Universidad de Costa Rica, San Jos\'{e}, Costa Rica 11501,}\\
		$^{2}${\it Escuela de F\'{i}sica, Universidad de Costa Rica, San Jos\'{e}, Costa Rica 11501,} \\
		$^{3}${\it Department of Materials Science and Metallurgy, University of Cambridge, Cambridge, UK CB3 0DS,} \\
		$^{4}${\it James Franck Institute, University of Chicago, 929 E 57 St, Chicago, Illinois, USA 60637,}\\
		$^{5}${\it Pritzker School of Molecular Engineering, University of Chicago, 5640 S Ellis Ave, Chicago, Illinois, USA 60637.}
		\end{center}

\vspace{0.5cm}
\begin{center}
	\textbf{Contents} \\
	\begin{itemize}
	    \item[] {\textbf{Note 1. Bare hybrid modes.}} 
	    \item[] {\textbf{Note 2. Phase stability.}} 
	    \item[] {\textbf{Note 3. Self-consistent phonon approximation.}} 
		\item[] {\textbf{Note 4. Fits to experiments in STO and KTO.}} 
	    \item[] {\textbf{References.}}	
	\end{itemize}
\end{center}

\setcounter{equation}{0}
\setcounter{figure}{0}
\setcounter{table}{0}
\setcounter{page}{1}
\makeatletter
\renewcommand{\theequation}{S\arabic{equation}}
\renewcommand{\thefigure}{S\arabic{figure}}
\renewcommand{\thetable}{S\arabic{table}}
\renewcommand{\bibnumfmt}[1]{[S#1]}
\renewcommand{\citenumfont}[1]{S#1}

\newpage

\section*{\label{sec:SN1} Supplementary Note 1. Bare hybrid modes}
In the absence of non-linearities ($u=0$),
we rewrite the GL Hamiltonian of Eq.~(1) in momentum space and in terms of the displacement field,
\begin{multline}
    \label{eq:Hq}
    H = \fint \frac{d^3 q}{(2\pi)^3} \, \sum_{\alpha \beta} U_{\alpha \beta} ({\bm q})  P_\alpha ({\bm q}) P^*_\beta ({\bm q})  +  \sum_{\alpha \beta} M_{\alpha \beta} ({\bm q})   u_\alpha ({\bm q}) u^*_\beta ({\bm q}) \\
    +  \sum_{\alpha \beta} V_{\alpha \beta} ({\bm q}) \left[ u_\alpha({\bm q}) P_\beta^*({\bm q}) + u_\alpha^*({\bm q}) P_\beta({\bm q})\right],
\end{multline}
where $ M_{\alpha \beta}({\bm q}) = \sum_{\gamma \lambda} C_{\gamma \alpha \lambda \beta } q_{\gamma} q_{\lambda},$ is the bare dynamical matrix of the acoustic phonons,
and $ V_{\alpha \beta}({\bm q}) = \sum_{\gamma \lambda} f_{\alpha \beta \gamma \lambda} q_{\gamma} q_{\lambda}$ is the flexoelectric interaction.  $\fint d^3 q$ indicates to integrate over half reciprocal space only. 

For isotropic media~\cite{SAharony1973a, SOliver2017a}, 
\begin{align*}
	U_{\alpha \beta} \left( {\bm q} \right) = \left[r + c q^2  \right] \delta_{\alpha \beta} + g_0 \left( q_{\alpha} q_{\beta} / q^2 \right) - h_0 q_{\alpha} q_{\beta},
\end{align*}
\begin{align*}
    C_{\alpha \beta \gamma \lambda} = \left( C_{v} - 2 C_s/3  \right) \delta_{\alpha \beta}  \delta_{\gamma \lambda} + C_s \left( \delta_{\alpha \gamma} \delta_{\beta \lambda} + \delta_{\alpha \lambda} \delta_{\beta \gamma} \right),
\end{align*}
and~\cite{SQuang2011a},
\begin{align*}
    f_{\alpha \beta \gamma \lambda} = \left( f_{v} - 2 f_s/3  \right) \delta_{\alpha \lambda}  \delta_{\beta \gamma} + f_s \left( \delta_{\alpha \gamma} \delta_{\beta \lambda} + \delta_{\alpha \beta} \delta_{\gamma \lambda} \right),
\end{align*}
where $C_{v}=(C_{11}+2C_{12})/3$ and $C_{s}=(C_{11}-C_{12})/2$ are the bulk and shear moduli and $f_{v}=(f_{11}+2f_{12})/3$ and $f_{s}=(f_{11}-f_{12})/2$ are bulk and shear components of the flexocoupling tensor. 

Thus,
\begin{align*}
    M_{\alpha \beta} ({\bm q}) = \left( C_v+C_s/3 \right) q_\alpha q_\beta + C_s q^2  \delta_{\alpha \beta},
\end{align*}
and,
\begin{align*}
    V_{\alpha \beta} ({\bm q}) = \left( f_v+f_s/3 \right) q_\alpha q_\beta + f_s q^2  \delta_{\alpha \beta},
\end{align*}
In the normal coordinates of the bare modes without flexoelectricity, the inverse propagator ${\bm G}_0^{-1}({\bm q})$ of Eq.~(\ref{eq:Hq}) is block-diagonal,
\begin{align}
    \label{eq:Goneloop}
    {\bm G}_0^{-1}({\bm q})=
    \begin{pmatrix}
       \omega_\text{TO}^2({\bm q}) & V_t({\bm q}) & 0 & 0 & 0 & 0 \\
       V_t({\bm q}) & \omega_\text{TA}^2({\bm q})  & 0 &  0  & 0 & 0 \\
       0 & 0 & \omega_\text{TO}^2({\bm q}) & V_t({\bm q}) & 0 & 0 \\
       0 & 0 & V_t({\bm q})  & \omega_\text{TA}^2({\bm q}) & 0 & 0\\ 
       0 &  0  & 0 & 0  & \omega_\text{LO}^2({\bm q}) & V_l({\bm q}) \\
       0 & 0 &  0 & 0 & V_l({\bm q}) &  \omega_\text{LA}^2({\bm q})
    \end{pmatrix},
\end{align}
where,
\begin{align*}
    \omega_\text{TO}^2({\bm q}) &= r + c q^2 + h q^4, 
    & 
    \omega_\text{LO}^2({\bm q}) &= r +g_0+ (c+h_0)q^2 , \\
    \omega_\text{TA}^2({\bm q}) &= a_t q^2,  &
    \omega_\text{LA}^2({\bm q}) &= a_l q^2,  \\
   V_t({\bm q}) &= v_t q^2,  &
   V_l({\bm q}) &= v_l q^2,
\end{align*}
with $a_t \equiv \hbar^2 C_s / \rho $, $a_l \equiv   \hbar^2 ( C_v + (4/3)C_s) / \rho $, $v_t  \equiv \hbar  f_s / \sqrt{\rho}$, and $v_l \equiv \hbar (f_v + (4/3)f_s ) / \sqrt{\rho}$, where $\rho$ is the density in units of mass per unit volume.  In writing Eq.~(\ref{eq:Goneloop}), we have chosen the following ordering for the entries  of  ${\bm G}_0^{-1}({\bm q})$: $\left\{ P_\text{T1}({\bm q}), u_\text{T1}({\bm q}), P_\text{T2}({\bm q}), u_\text{T2}({\bm q}), P_\text{L}({\bm q}), u_\text{L}({\bm q}) \right\}$, where T and L refer to transverse and longitudinal components relative to ${\bm q}$, respectively. Note that $V_{\alpha \beta}({\bm q})$ does not mix the transverse and longitudinal modes and that the transverse blocks are identical. These are consequences of assuming an isotropic medium and do not generally occur in crystals. Note we have added a $\mathcal{O}(q^4)$ term in $\omega_\text{TO}^2({\bm q})$ by hand. 
This is done to produce a well-conditioned hybrid TA branch for $f_s > f_{cr}$ that would allow us to capture the modulated wavevector in the hybrid modes (otherwise it would be negatively unbounded, which is unphysical). 
We note such terms can result from higher-order gradients in the effective Hamiltonian (e.g.,  $\norm{\nabla {\bm P}({\bm x})}^4$). We do not explicitly write them down in Eq.~(1)  
for simplicity, as it is sometimes done in the continuum theory of incommensurate transitions~\cite{SStrukov1998a}. We also note that a similar term could be added to $\omega_\text{TA}^2({\bm q})$, however, we do not do so as it that would only add an extra fitting parameter with no qualitative difference in the resulting hybrid modes.

Since $G_0^{-1}({\bm q})$ is a dynamical matrix, the bare frequencies are given by its eigenvalues,
\begin{subequations}
   \label{eq:hybridmodes}
\begin{align}
    \Omega_{\text{TO}}^0({\bm q}) &=\frac{1}{\sqrt{2}}\left(   \omega_\text{TA}^2({\bm q}) +\omega_\text{TO}^2({\bm q}) +  \sqrt{4V_t^2({\bm q})+ \left[ \omega_\text{TO}^2({\bm q}) -\omega_\text{TA}^2({\bm q}) \right]^2}\right)^{1/2}, \\
    \Omega_\text{TA}^0({\bm q})  &= \frac{1}{\sqrt{2}} \left( \omega_\text{TA}^2({\bm q}) +\omega_\text{TO}^2({\bm q}) -  \sqrt{4V_t^2({\bm q})+ \left[ \omega_\text{TO}^2({\bm q}) -\omega_\text{TA}^2({\bm q}) \right]^2} \right)^{1/2}, \\
    \Omega_\text{LO}^0({\bm q}) &=  \frac{1}{\sqrt{2}} \left( \omega_\text{LA}^2({\bm q}) +\omega_\text{LO}^2({\bm q}) +  \sqrt{4V_l^2({\bm q})+ \left[ \omega_\text{LO}^2({\bm q}) -\omega_\text{LA}^2({\bm q}) \right]^2}\right)^{1/2}, \\
    \Omega_\text{LA}^0({\bm q})  &= \frac{1}{\sqrt{2}} \left( \omega_\text{LA}^2({\bm q}) +\omega_\text{LO}^2({\bm q}) -  \sqrt{4V_l^2({\bm q})+ \left[ \omega_\text{LO}^2({\bm q}) -\omega_\text{LA}^2({\bm q}) \right]^2}\right)^{1/2},
\end{align}  
\end{subequations}
where the transverse and longitudinal branches are two- and one-fold degenerate, respectively.  The TA, TO, LA, and LO labels refer to optic- and acoustic-like mode behavior in the long-wavelength limit. %We note Eqs.~(\ref{eq:hybridmodes}) are only approximate as we have ignored the dynamic component of the flexoelectric effect~\cite{SKvasov2015a}.

We now discuss the behavior of the hybrid modes. For $0 < v_t   < v_{cr} \equiv \sqrt{c \, a_t}$, the TA an TO phonons repel at finite wavevectors and soften as $T \to T_0$. At $T_0$, $\Omega_\text{TO}^0({\bm q}=0)$ condenses, which leads to a FE transition. For $v_t \geq v_{cr}$, the level repulsion generates a minimum at a finite wavevector $q_{mod}$ in the acoustic branch as temperature decreases. At a temperature $T_{mod}=T_0 + (v_t^2-v_{cr}^2)^2/(4h r_0 a_t^2 )$, $q_{mod}=\sqrt{(v_t^2-v_{cr}^2)/(2 h a_t)}$ and 
$\Omega_\text{TA}^0({\bm q}_{mod})$ condenses, while the zone-center TO phonon remains gaped with  energy $\tau^0_{mod} \equiv \Omega_\text{TO}^0({\bm q}=0)=\sqrt{r_0(T_{mod}-T_0)}=\left(v_{cr}^2-v_t^2\right)/(2 a_t \sqrt{h})$. We note that $v_t=v_{cr}$ is a triple point between the paraelectric, FE and modulated phases. N.B.: since $f_s / f_{cr} = v_t/v_{cr} $, where 
$f_{cr} = \sqrt{c \, C_s}$, the above results can be rewritten as follows: $T_{mod}=T_0 + (f_s^2-f_{cr}^2)^2/(4hr_0 C_s^2)$, $q_{mod}=\sqrt{(f_s^2-f_{cr}^2)/(2h C_s)}$, $\Omega_\text{TO}^0({\bm q}=0)=\left(f_{cr}^2-f_s^2\right)/(2C_s\sqrt{h})$. These resulting phase diagram is shown in Fig.~1\,(a).

For the longitudinal modes, the relevant parameter is $v_l$: for $0 \leq v_l < v_{l,cr} \equiv \sqrt{\left(c+h_0\right)\,a_l}$, the zone center LO phonon does not condense due to the depolarizing field $g_0$.  
For $  v_l \geq v_{l,cr}$, $\Omega_\text{LA}^0({\bm q}_{l,mod})$ condenses at a temperature $T_{l,mod}=T_0 - (g_0/r_0) + \left(v_l^2-{v_{l,cr}}^2\right)^2/(4h a_l^2 r_0)$, where $q_{l, mod} = \sqrt{\left(v_l^2-v_{l,cr}^2\right)/(2 a_l h)}$. The zone-center LO phonon remains gaped with (squared) energy $\Omega_\text{LO}^0({\bm q}=0)= \left(v_{l,cr}^2-{v_l}^2\right)/(2a_l \sqrt{h}) $.  We note that $T_{mod}$ and $T_{l,mod}$ depend for the most part on independent model parameters, therefore one cannot
conclude in general which modulated transition precedes the other. Nonetheless, $g_0$ is typically very large in pervoskite FEs, thus we expect $T_{mod} > T_{l,mod}$, which will be assumed hereafter. 

\newpage 

\section*{\label{sec:SN2} Supplementary Note 2. Phase stability}

We now discuss the stability of the PB. We will show that in the presence of purely thermal fluctuations the  FE PB is stable, while the modulated solid and triple point in the phase diagram are unstable. 

At the Gaussian level and in the classical limit, the local correlation functions $\bra \norm{{\bm P}^2} \ket$ of the disordered phase of our isotropic model are given in terms of its transverse ($\bra P_\perp^2({\bm q})\ket_0 \equiv \bra P_\text{ T1}^2({\bm q}) \ket_0 = \bra P_\text{T2}^2({\bm q}) \ket_0$) and longitudinal fluctuations ($\bra P_\parallel^2({\bm q})\ket_0 \equiv \bra P_\text{L}^2({\bm q}) \ket_0 $) as follows,
\begin{align}
    \bra \norm{{\bm P}^2} \ket_0 =  \int \frac{d^3 q}{(2\pi)^3} \left[  2 \bra P_\perp^2({\bm q})\ket_0 + \bra P_\parallel^2({\bm q})\ket_0  \right],
\end{align}
where,
\begin{align*}
     \bra P_\perp^2({\bm q})\ket_0  
    = \frac{k_B T}{r_0(T-T_0)+ (1-v_t^2/v_{cr}^2) c q^2 + h q^4},
\end{align*}
and,
\begin{align*}
   \bra P_\parallel^2({\bm q})\ket_0 
    = \frac{k_B T}{r_0(T-T_0)+ g_0 + (1-v_l^2/v_{l,cr}^2) (c+h_0)  q^2 + h q^4}.
\end{align*}
At the onset of the FE transition ($T \to T_0$ for $0\leq v_t < v_{cr}$),  $ \bra P_\perp^2({\bm q})\ket_0 \to \infty$ as $q \to 0$, while  $\bra P_\parallel^2({\bm q}) \ket_0$ remains finite due to the depolarizing field term $g_0$ and therefore we will ignored it. By integrating $\bra P_\perp^2({\bm q})\ket_0$ over all of ${\bm q}$-space, we obtain the following result for $T \to T_0$,
\begin{align*}
   (k_B T)^{-1} \int d^3 q \bra P_\perp^2({\bm q})\ket_0   &\propto \frac{\pi}{2\sqrt{\left( 1-v_t^2/v_{cr}^2 \right) h c}} \left[1 
   - \sqrt{\frac{h}{c}\frac{\left(T-T_0\right)}{\left( 1-v_t^2/v_{cr}^2 \right)}} \right].
\end{align*}
The thermal fluctuations are finite at $T_0$, thus  
the long-range FE order is stable in the purely classical limit. 

We now look at the fluctuations near above the modulated transition.  For $T \to T_{mod}$ and $v_t \geq v_{cr}$,  $\bra P_\perp^2({\bm q}) \ket_0 \to \infty$ as $q \to q_{mod}$, whereas  $\bra P_\parallel^2({\bm q}) \ket_0$ remains finite since $T_{mod} > T_{l, mod}$ by assumption and thus we ignore it. By integrating the transverse fluctuations over momentum, we obtain the following result for $T \to T_{mod}$,
\begin{align}
    \label{eq:flucMod}
   (k_B T)^{-1} \int d^3 q \bra P_\perp^2({\bm q}) \ket_0 \propto  \left(\frac{T_{mod}-T_0}{T-T_{mod}}\right)^{1/2}. 
\end{align}
Equations~(\ref{eq:flucMod})  diverges as $T \to T_{mod}$, thus the modulated solid is unstable. 

At the triple point ($v_t=v_{cr}$), $T_{mod}=T_0$ and $q_{mod}=0$, thus,
\begin{align}
    \label{eq:flucCoex}
   (k_B T)^{-1} \int d^3 q  \bra P_\perp^2({\bm q}) \ket_0 \propto  \left(\frac{1}{T-T_0}\right)^{1/4}. 
\end{align}

Equation (\ref{eq:flucCoex}) diverges as $T \to T_0$, therefore the coexistence is unstable. 

\newpage

\section*{\label{sec:SN3} Supplementary Note 3. Self-consistent phonon approximation}
\vspace{-1cm}
We follow a standard procedure~\cite{SChaikin1995a}
to solve the statistical mechanical problem posed by our model Hamiltonian in disordered phase. 
At the Gaussian level, the quartic anharmonicities in  Eq.~(1) are approximated by,
\begin{align*}
    \int d^3 x \sum_{\alpha ,\beta} u  P^2_\alpha({\bm x}) P^2_\beta({\bm x}) 
    \simeq   \int d^3 x \sum_{\alpha, \beta } \left[ 2 u \bra \norm{{\bm P}({\bm x})}^2 \ket \delta_{\alpha \beta}  + 4 u \bra P_\alpha({\bm x}) P_\beta({\bm x}) \ket \right] P_\alpha({\bm x})  P_\beta({\bm x}),
\end{align*}
where $\bra P_\alpha({\bm x}) P_\beta({\bm x}) \ket$ are correlation functions of polarization and $\bra \norm{{\bm P}({\bm x})}^2 \ket = \sum_{\gamma=1}^3  \bra P^2_\gamma({\bm x}) \ket$. We thus have an effective Hamiltonian identical to Eq.~(1), but with $U_{\alpha \beta}({\bm x},{\bm x}^\prime)$ now given as follows, 
\begin{align*}
   U_{\alpha \beta}({\bm x},{\bm x}^\prime) = \left[ \left( r +  4 u \bra \norm{{\bm P}({\bm x})}^2 \ket \right) \delta_{\alpha \beta} + 8 u \bra P_\alpha({\bm x}) P_\beta({\bm x}) \ket \right] \delta({\bm x}-{\bm x}^\prime)   
    + F_{\alpha \beta}({\bm x},{\bm x}^\prime).
\end{align*}
By symmetry, $\bra P_1({\bm x}) P_2({\bm x}) \ket = \bra P_1({\bm x}) P_3({\bm x}) \ket = \bra P_2({\bm x}) P_3({\bm x}) \ket = 0$ and $\bra P_1^2({\bm x}) \ket = \bra P_2^2({\bm x}) \ket = \bra P_3^2({\bm x}) \ket = (1/3) \bra \norm{\bm P}^2\ket $, thus, 
\begin{align}
    \label{eq:dressedU}
    & U_{\alpha \beta}({\bm x},{\bm x}^\prime) = \left[r +   (20u/3) \bra \norm{\bm P}^2\ket  \right] \delta_{\alpha \beta} \delta({\bm x}-{\bm x}^\prime) 
    + F_{\alpha \beta}({\bm x},{\bm x}^\prime). 
\end{align}
The dressed propagator ${\bm G}^{-1}({\bm q})$ and frequencies $\Omega_\lambda({\bm q}),~(\lambda =\text{TO, TA, LO, LA})$ are therefore identical to their corresponding bare forms given in Eqs.~(\ref{eq:Goneloop}) and~(\ref{eq:hybridmodes}) with $r$ replaced by,  
\begin{align}
    \label{eq:TOSM}
    \tau \equiv \Omega_\text{TO}^2({\bm q}=0) = r +   (20u/3) \bra \norm{\bm P}^2\ket,
\end{align}
where $\Omega_\text{TO}(0)$ is the zone-center TO mode frequency. The correlation functions are calculated from the fluctuation-dissipation theorem~\cite{SPytte1970a},
\begin{align}
        \label{eq:correlations2}
        \bra P_\alpha ({\bm x}) P_\beta  ({\bm x}^\prime)  \ket =  \int \frac{d^3 q}{(2\pi)^3} e^{i {\bm q}\cdot \left({\bm x}- {\bm x}^\prime\right)} \sum_{\lambda \lambda^\prime } b_{\alpha \lambda}({\bm q}) \psi_{\lambda \lambda^\prime}({\bm q})  b_{\lambda^\prime \beta}^\dagger({\bm q}),
\end{align}
where, 
\begin{align*}
        \psi_{\lambda \lambda^\prime}({\bm q}) =  \frac{1}{2 \Omega_\lambda({\bm q})} \coth\left( \frac{  \Omega_\lambda({\bm q})}{2 k_B T}\right) \, \delta_{\lambda \lambda^\prime},
\end{align*}
and $b_{\alpha \lambda}({\bm q})$ is a unitary transformation that diagonalizes ${\bm G}^{-1}({\bm q})$. Eq.~(\ref{eq:correlations2}) provides the self-consistency equation for $\Omega_\lambda({\bm q})$.

\subsection{Dependence of \texorpdfstring{$\tau$}{TEXT} on \texorpdfstring{$u$}{TEXT} for \texorpdfstring{$f_s>f_{cr}$}{TEXT}}

\begin{figure}[h!]
    \centering
    \includegraphics[scale=0.45]{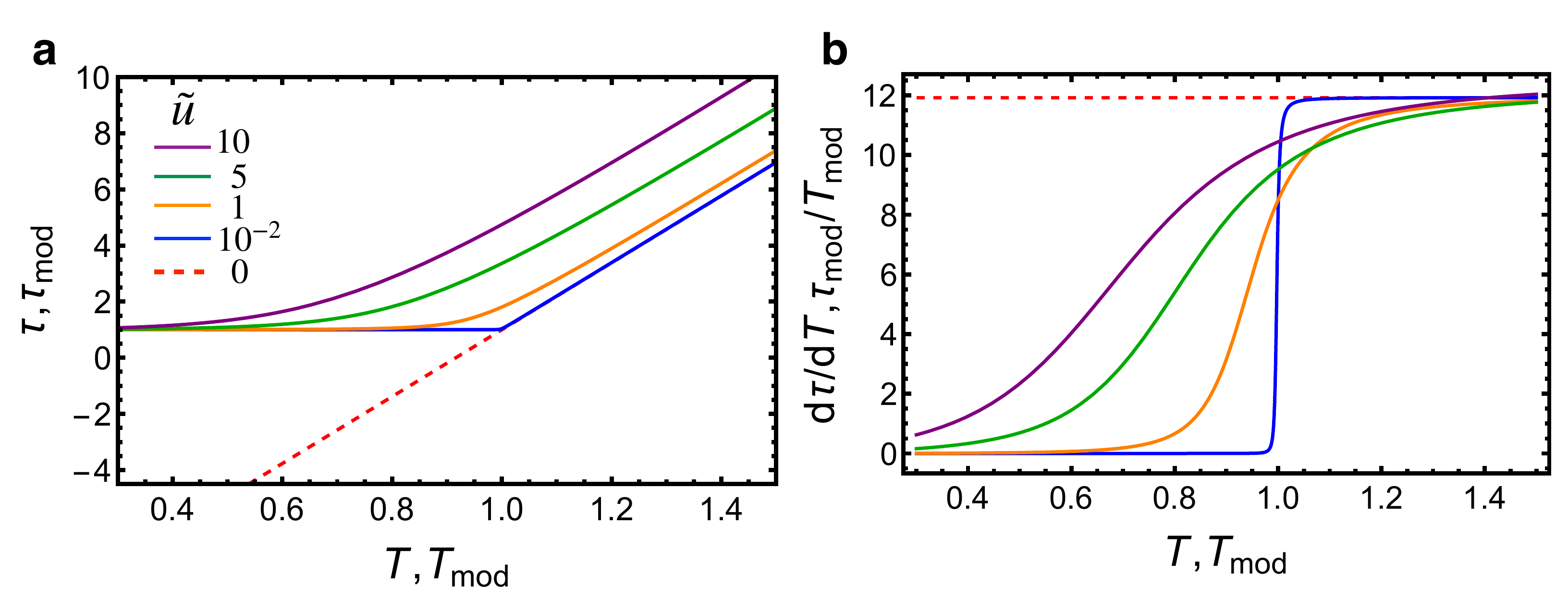}
    \caption{(a) Temperature dependence of the zone-center TO mode $\tau$ and (b) its first derivative for several values of the dimensionaless quartic coefficient $\tilde{u} \equiv (20/3 \pi) (u q_{mod}^3/\sqrt{\tau_{mod}})$. All other parameters are taken from Table I of the main text for STO.}
    \label{fig:kink}
\end{figure}

Figs.~\ref{fig:kink}~(a) and (b) above show the temperature dependence of  $\tau \equiv \Omega_\text{TO}^2(0)$ and its first derivative $d\tau/dT$ for several values of $u$. 
We first point out the dramatic change in the behavior of $\tau$ between $u=0$ and $u>0$ is expected as infinitesimally small fluctuations prevent long-range modulated order. We then note that for finite $u$, $d\tau/dT$ seems to become multi-valued at $T_{mod}$ as $u \to 0$ (but remains finite). However,  a closer inspection reveals that it is well-defined. 

For large flexocouplings ($f_s > f_{cr}$), we  rewrite Eq.~(\ref{eq:TOSM}) as follows,
\begin{align}
    \label{eq:tau1}
    \tau -\tau_{mod}^0 = r_0\left(T-T_{mod}\right) + \left(20u/3\right) \bra \norm{{\bm P}(0)}^2 \ket,
\end{align}
where $\bra \norm{{\bm P}(0)}^2 \ket$ are the local correlation functions of polarization, and 
$\tau_{mod}^0 \equiv r_0 \left(T_{mod}-T_0\right)$ is the harmonic ($u=0$) zone-center TO frequency at $T_{mod}$.

In  the classical limit and $\tau - \tau_{mod}^0 \to 0$, %
\begin{align*}
    \bra \norm{{\bm P}(0)}^2 \ket = \int \frac{d^3 q}{(2\pi)^3} \frac{k_B T}{\left(\tau - \tau_{mod}^0\right) + h\left(q^2 - q_{mod}^2\right)^2} \propto \frac{k_B T}{\left(\tau -\tau_{mod}^0 \right)^{1/2}}.
\end{align*}
We now substitute this into Eq.~(\ref{eq:tau1}), 
\begin{align*}
    %\label{eq:tau2}
    \tau-\tau_{mod}^0 \simeq r_0\left(T-T_{mod}\right) + \frac{u^\prime T}{\left(\tau -\tau_{mod}^0 \right)^{1/2}},
\end{align*}
with $u^\prime \propto u$. By taking the derivative with respect to $T$ on both sides of this equation and solve for $d \tau / dT$, we obtain the following result,
\begin{align*}
    \frac{d \tau}{ dT}  = \frac{r_0+u^\prime / \left(\tau -\tau_{mod}^0 \right)^{1/2}}{1+u^\prime T / 2\left(\tau -\tau_{mod}^0 \right)^{3/2}}.
\end{align*}
For the harmonic theory (i.e. $u^\prime = u = 0$), we get $d\tau /dT = r_0$ at all temperatures, as expected. For $u^\prime >0$ and $T=T_{mod}$, $\tau -\tau_{mod}^0 \simeq \left(u^\prime T_{mod}\right)^{2/3}$ and,
\begin{align*}
    \left.  \frac{d \tau}{dT} \right|_{T=T_{mod}}  = \frac{2}{3}\left(r_0+\frac{{u^\prime}^{2/3}}{T_{mod}^{1/3}}\right).
\end{align*}

We thus find $d \tau / dT$ is single-valued at $T_{mod}$ for $u^\prime >0$ and thus there is no kink.

\subsection{Correlation function of polarization in the lamellar-like liquid phase}

The purpose of this subsection is the derive the correlation function of polarization given in Eq.~(2) of the main text. We consider the  correlation function of polarization given as follows,
\begin{align}
    \label{eq:Scorrelation1}
    \bra {\bm P}({\bm x} ) \cdot  {\bm P}({\bm 0} ) \ket &= \sum_\alpha \bra P_\alpha({\bm x} ) P_\alpha({\bm 0} ) \ket 
    =   \int \frac{d^3 q}{(2\pi)^3} e^{i {\bm q}\cdot {\bm x}}  \left[  2 \bra P_\perp^2({\bm q})\ket + \bra P_\parallel^2({\bm q})\ket  \right],
\end{align}
where $\bra P_\perp^2({\bm q})\ket$ and $\bra P_\parallel^2({\bm q})\ket$ are transverse and longitudinal polarization correlations. In the classical limit,
\begin{align}
    \label{eq:Scorrelation2}
     \bra P_\perp^2({\bm q})\ket  
    &= \frac{k_B T}{\tau + (1-f_s^2/f_{cr}^2) c q^2 + h q^4},
\end{align}
and, 
\begin{align*}
   \bra P_\parallel^2({\bm q})\ket 
    = \frac{k_B T}{\tau + g_0 + (1-f_l^2/f_{l,cr}^2) (c+h_0)  q^2 + h q^4}.
\end{align*}
where  $\tau$ is given by Eq.~(\ref{eq:TOSM}) and  $q_{mod}$ and $\tau_{mod}^0$ are given in SN1. Since the longitudinal fluctuations are much weaker than the transverse ones due to the large depolarizing term $g_0$, we will ignore them. 

For $f_s> f_{cr}$, we rewrite Eq.~(\ref{eq:Scorrelation2}) as follows,
\begin{align*}
     \bra P_\perp^2({\bm q})\ket  
     = \frac{q_{mod}^4}{\tau_{mod}^0} \frac{k_B T}{ \xi^{-4} +(q^2-q_{mod}^2)^2},
\end{align*}
where $\xi$ is the correlation length,
\begin{align*}
    q_{mod} \, \xi \equiv \left(\frac{\tau^0_{mod}}{\tau - \tau^0_{mod}}\right)^{1/4}.
\end{align*}
Since the fluctuations are isotropic, the angular integrals in Eq.~(\ref{eq:Scorrelation1}) can be readily calculated,
\begin{align*}
     \bra {\bm P}({\bm x} ) \cdot  {\bm P}({\bm 0} ) \ket 
     = \frac{k_B T}{\pi^2} \frac{q_{mod}^4}{ \tau_{mod}^0}  \int_0^\infty dq \,  \frac{q^2}{ \xi^{-4} +(q^2-q_{mod}^2)^2} \frac{\sin q x  }{q  x },
\end{align*}
where $x=\norm{\bm x}$. After evaluating this integral by countour integration and the residue theorem, we obtain the following result,
\begin{align*}
    \bra {\bm P}({\bm x} ) \cdot  {\bm P}({\bm 0} ) \ket &= \frac{k_B T}{2\pi \xi^{-2} } \frac{q_{mod}^5}{ \tau_{mod}^0} e^{- q_{mod} \, r \sqrt{\frac{\sqrt{1+(q_{mod}\xi)^{-4}}-1}{2}}} \nonumber \\
    &~~~~~~~~~~~~~~~~ 
    \times 
    \sin \left( q_{mod} \, r \sqrt{\frac{\sqrt{1+(q_{mod}\xi)^{-4}}+1}{2}} \right) / (q_{mod} x).
\end{align*}
For large correlation lengths ($q_{mod}\, \xi \gg 1$), we obtain,
\begin{align*}
    \bra {\bm P}({\bm x}) \cdot {\bm P}({\bm 0})  \ket \simeq  \frac{k_B T}{2\pi \xi^{-2}} \frac{q_{mod}^5  }{ \tau_{mod}^0} \, e^{-x/(2\,  q_{mod} \, \xi^2)} \,  \sin \left( q_{mod} \, x \right)/\left( q_{mod} \, x \right),
\end{align*}
which is the desired result. We note the calculation of the correlation function in the Gaussian approximation is identical to the one presented here except that one  must replace $\tau \to r_0(T-T_0)$ in Eq.~(\ref{eq:Scorrelation2}) and define the correlation length as $q_{mod} \, \xi = \left(T_{mod}/(T-T_{mod})\right)^{1/4}$. 

\newpage

\section{\label{sec:SN4} Supplementary Note 4. Fits to experiments in STO and KTO}

In this note, we describe our fits to the phonon spectra of STO and KTO. Our model has seven parameters $r_0, T_0, u, c, a_t, v_t $ and $h$, which are given in Table~I of the main text. For both materials, we have attempted fits for $f_s \geq f_{cr}$ and $0 \leq f_s < f_{cr}$. 

\vspace{-0.75cm}

\subsection{Large couplings: \texorpdfstring{$f_s \geq f_{cr}$}{TEXT} (i.e.  \texorpdfstring{$v_t \geq v_{cr}$}{TEXT})}
\vspace{-0.25cm}
\underline{STO:} We take $r_0, T_0$ and $u$ from Landau theory~\cite{SModernPerspective2007a}. Values for $c, a_t$, and  $v_t$ are available in Ref.~\cite{SVaks1973a},  however, they do not provide a good fit to the recent neutron scattering experiments of Ref.~\cite{SFauque2022a}. We thus provide a new parametrization. 

From the Supplementary Note 2, we have $a_t = \hbar^2 C_s / \rho$. We take $C_s=1.24\times 10^{11}$\,N m$^{-2}$ from the observed low-temperature shear modulus and $\rho=5.104\times 10^3\,$kg m$^{-3}$\,\cite{SCarpenter2007a}. To obtain $c$, we  expand Eq.~(\ref{eq:hybridmodes}a) to leading order in $q$, i.e., $\Omega_\text{TO}^2({\bm q}) \simeq \tau + c q^2$ and fit it to the observed TO mode dispersion at 10\,K~\cite{SFauque2022a}. $v_t$ and $h$ were obtained by fitting $\tau_{mod}^{1/2}$ and $q_{mod}$ to the zone center TO energy at low temperatures ($\simeq  1.70\,\text{meV}$)~\cite{SYamada1969a, SVogt1995a} and to the wavevector minimum in the TA branch ($\simeq  0.024\,$rlu)~\cite{SFauque2022a}, respectively.  

\underline{KTO:} To determine $r_0$, $T_0$, and $u$, we first take the Landau limit of our model with a homogeneous polarization (${\bm P}({\bm x}) = \bra  {\bm P} \ket_\text{MFT}$) and $f_{\alpha \beta \gamma \lambda}=F_{\alpha \beta}({\bm x}, {\bm x}^\prime)=0$. We then fit $r_0$, $T_0$, and $u$ to the observed temperature and applied electric field dependence of the dielectric constant~\cite{SFujishita2016a, SSkoromets2016a}.
The values for $c$ and  $a_t$ are known from neutron scattering experiments Ref.~\cite{SFarhi2000a}. $v_t$ and $h$ are  obtained by fitting $\tau_{mod}^{1/2}$ and $q_{mod}$ to the observed zone center TO energy at low temperatures ($\simeq 2.5\,\text{meV}$)~\cite{SShirane1967a,SVogt1995a,  SFarhi2000a} and the wavevector minimum in the TA branch $(\simeq 0.1\,$rlu)~\cite{SAxe1970a}, respectively.  

\vspace{-0.75cm}

\subsection{Small couplings: \texorpdfstring{$0 \leq f_s < f_{cr}$}{TEXT} (i.e.  \texorpdfstring{$0 \leq v_t < v_{cr}$}{TEXT})}

\vspace{-0.25cm}

\underline{STO and KTO:} We assume there is no long-range polar order in STO and KTO down to absolute zero, i.e., the energy gap of the zone-center TO phonon is finite at $0\,$K (see Eq.~(\ref{eq:TOSM})),
\begin{align}
    \label{eq:TOgap}
    \tau=\Omega_\text{TO}^2({\bm q}=0) = -r_0 T_0 + \left( 20u/3 \right)  \bra \norm{\bm P}^2 \ket > 0 ,
\end{align}
where $\bra \norm{\bm P}^2 \ket$ are the fluctuations of polarization calculated within the SCPA (see Supplementary Note 3). By taking $r_0, T_0, u, c, a_t$ and $h$ from the previous section and setting $\tau = 1.7\,\text{meV}\,( 2.5\,\text{meV})$ for STO (KTO), we find that the inequality (\ref{eq:TOgap}) cannot be satisfied for $ 0 \leq v_t < v_{cr}$ in either material.  We find the same result if we attempt to fit $r_0, T_0$, and $u$ while keeping the polarization physical, which we did by constraining $r_0, T_0$, and $u$ to produce the Landau value at $0\,$K, i.e., 
 $\bra {\bm P} \ket_\text{MFT}\, = \left(\sqrt{ 2 \epsilon_0} A  \times 10^2\right) \sqrt{r_0 T_0/(2u)}= 8.6\,$$\mu$C\,/cm$^2\,(2.3\,$$\mu$C\,/cm$^2)$ for STO (KTO), where $\epsilon_0$ is the vacuum permittivity in units of C$^2\,$N$^{-1}\,$m$^{-2}$, and $A=194.4\,$meV\,($158.7\,$meV) is a proportionality factor between the experimental zone-center TO frequency ($\Omega_\text{TO}^\text{exp}$) and the dielectric constant $(\epsilon^\text{exp})$:  $\Omega_\text{TO}^\text{exp}=A / \sqrt{\epsilon^\text{exp}}$~\cite{SYamada1969a}.  $\sqrt{ 2 \epsilon_0} A  \times 10^2 $ is a conversion factor from our model units to $\mu$C\,/cm$^2$. 

\begin{figure}[htp!]
    \centering
    \includegraphics[scale=0.6]{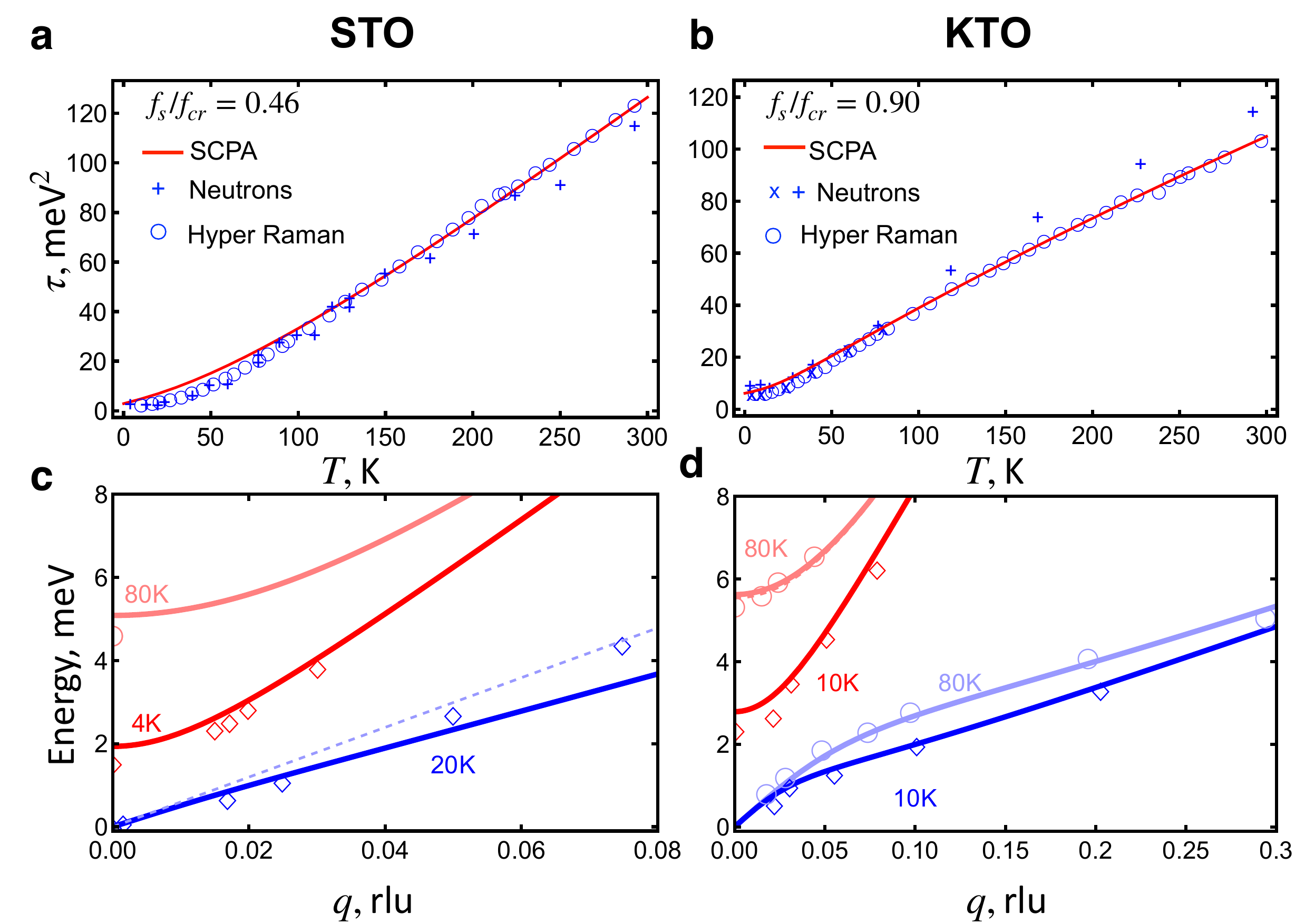}
    \caption{{\bf (a),(b)} Calculated temperature dependence of the zone center hybrid TO mode compared to neutron~\cite{SShirane1967a, SYamada1969a, SFarhi2000a} and hyper-Raman~\cite{SVogt1995a} scattering experiments . {\bf (c),(d)} Calculated dispersion of the hybrid TA and TO phonons (solid lines) compared to neutron scattering experiments (circles and diamonds)~\cite{SFauque2022a, SFarhi2000a}.  Dashed line in (d) is the bare linear TA dispersion for $f_s=0$ with a sound velocity of 4800\,m/s (as shown in Ref.~\cite{SFauque2022a} for comparison). Model parameters are given in Table~\ref{St:parameters}.}
    \label{Sfig:comparison}
\end{figure}

By relaxing the polarization constraint and fitting $r_0, T_0$ and $u$ to the observed zone-center TO mode, we find a fair agreement with experiments, as shown in Fig.~\ref{Sfig:comparison}.  However, the resulting parametrization yields an unphysical $\bra {\bm P} \ket_\text{MFT}\,$ (see Table~\ref{St:parameters}).

\begin{table}[htp]
\caption{Model parameters for small flexocoplings and no polarization constrain. For STO, we took $c$ from  Table~I,  $v_t$ from Ref.~\cite{SVaks1973a}, and obtained $r_0, T_0$ and $u$ by fitting Eq.~(\ref{eq:TOSM}) to $\tau^{1/2}=1.7, 6.7$ and $11.1\,$meV at $T=0,127$, and $292\,$K, respectively, with $h=0$. For KTO, we took $c, a_t$ and $v_t$ from Ref. \cite{SFarhi2000a}, and obtained $r_0, T_0$ and $u$ by fitting Eq.~(\ref{eq:TOSM}) to $\tau^{1/2}=2.5, 7.1$ and $10.2\,$meV at $T=0,131$, and $297\,$K, respectively, with $h=0$. We could not fit the data for either material for $h>0$.} 
    \label{St:parameters}
   \begin{tabular}{l|r|r} 
   \hline
    &  \hspace{1cm} KTO &   \hspace{1cm} STO  \\ \hline \hline 
       $r_0\,$[meV$^2$ K$^{-1}$]  & $0.16$ & $0.20$    \\
       $T_0\,$[K]  & $100$ & $473$ \\
       $u\,$ [$10^{3}$ meV$^3$   rlu$^{-3}$]  & $0.12$  & $1.29$   \\
       $c\,$[$10^3$\, meV$^2$ rlu$^{-2}$]  & $4.828$\footnotemark[1]  &     $13.5$ \\
       $a_t$\, [$10^3\,$ meV$^2$  rlu$^{-2}$]  & $1.553$\footnotemark[1] &  $2.739$\footnotemark[2]   \\
     $v_t$\, [$10^3\,$ meV$^2$ rlu$^{-2}$] &  $2.451$\footnotemark[1]  & $2.802$\footnotemark[3]  \\
     $h\,$ [meV$^4$  rlu$^{-4}$]  & $0.0$  & $0.0$  \\
       \hline 
      $v_t/v_{cr} = f_s/f_{cr}$   & $0.90$    & $0.50$  \\ \hline
      $\bra {\bm P} \ket_\text{MFT}\,$ [$\mu$C/cm$^2$] & $0.02$ &  $0.02$ 
      \\ \hline 
   \end{tabular}
    \footnotetext[1]{From Ref.~\cite{SFarhi2000a}.}
    \footnotetext[2]{From Ref.~\cite{SCarpenter2007a}.}
    \footnotetext[3]{From Ref.~\cite{SVaks1973a}.}
\end{table}

\newpage

%\bibliography{referencesSM}

%

\end{document}